\documentclass[conference]{IEEEtran}

\usepackage{endnotes}
\usepackage{balance}
\usepackage{amsfonts}
\usepackage{epsfig}
\usepackage{amssymb}
\usepackage{amscd}
\usepackage{amsmath}
\usepackage{microtype}
\usepackage{array}
\usepackage{graphics}
\usepackage{epsfig}
\usepackage{makecell}

\usepackage{listings}
\usepackage{graphicx}
\usepackage{subcaption}
\usepackage{amsmath}
\usepackage{amsfonts}

\usepackage{listings}
\usepackage[hyphens]{url}
\usepackage{soul}
\usepackage{diagbox}

\usepackage{tabularx}
\usepackage{epstopdf}
\usepackage{threeparttable}
\usepackage{setspace}
\usepackage{lipsum}
\usepackage{color,xcolor}

\usepackage{cite}

\usepackage{forloop}

\newcommand{\ignore}[1]{}
\newcommand{\revised}[1]{}

\newcolumntype{M}[1]{>{\centering\arraybackslash}m{#1}}
\newcolumntype{C}[1]{>{\centering\let\newline\\\arraybackslash\hspace{0pt}}m{#1}}
\newcolumntype{L}[1]{>{\raggedright\let\newline\\\arraybackslash\hspace{0pt}}m{#1}}

\newcommand{\etal}{{\textit{et al.} }}

\newcommand{\ie}{\textit{i}.\textit{e}.}

\DeclareMathOperator*{\argmin}{arg\,min}
\newcommand\zztitle[1]{ \vspace{5pt}\noindent\textbf{#1} }

\begin{document}

\title{Invisible Mask: Practical Attacks on Face Recognition with Infrared}

\author{Zhe Zhou$^1$, Di Tang$^2$, Xiaofeng Wang$^3$, Weili Han$^1$, Xiangyu Liu$^4$, Kehuan Zhang$^2$\\
$^1$Fudan Univiersity, $^2$CUHK, $^3$IUB, $^4$Alibaba Inc.\\
$^1$zhouzhe@fudan.edu.cn
}

\maketitle

\begin{abstract}
Accurate face recognition techniques make a series of critical applications possible: policemen could employ it to retrieve criminals' faces from surveillance video streams; cross boarder travelers could pass a face authentication inspection line without the involvement of officers. Nonetheless, when public security heavily relies on such intelligent systems, the designers should deliberately consider the emerging attacks aiming at misleading those systems employing face recognition.

We propose a kind of brand new attack against face recognition systems, which is realized by illuminating the subject using infrared according to the adversarial examples worked out by our algorithm, thus face recognition systems can be bypassed or misled while simultaneously the infrared perturbations cannot be observed by raw eyes. Through launching this kind of attack, an attacker not only can dodge surveillance cameras. More importantly, he can impersonate his target victim and pass the face authentication system, if only the victim's photo is acquired by the attacker. Again, the attack is totally unobservable by nearby people, because not only  the light is invisible, but also the device we made to launch the attack is small enough. According to our study on a large dataset, attackers have a very high success rate with a over 70\% success rate for finding such an adversarial example that can be implemented by infrared. To the best of our knowledge, our work is the first one to shed light on the severity of threat resulted from infrared adversarial examples against face recognition.
\end{abstract}

\section{Introduction}
With the advances of deep learning (DL) and pattern recognition, face recognition (FR) systems powered by these technologies are evolving rapidly, becoming increasingly capable over years. Today, the state of the art FR system can already outperform humans in terms of recognition accuracy~\cite{sun2014deep,zhu2015high,taigman2014deepface,sun2015deepid3}. This enables its wide deployment in the real world, serving various applications such as attendance recording, device login, and even border inspection.


\vspace{5pt}\noindent\textbf{Challenges of real-world adversarial learning}.  Despite the great enhancement in accuracy, existing FR techniques, however, are actually not as reliable as one expects. Prior research shows that a picture, with only minor changes to its pixels, could fool the recognition system into misidentifying one subject as another~\cite{szegedy2013intriguing, carlini2017towards}. Such an attack, known as \textit{adversarial learning}, has been found to be hard to defend against~\cite{adv_exp_hard}, particularly when the adversary is capable of arbitrarily changing the input of the FR system, e.g., editing every pixel on a picture. In reality, however, he may  end up with only \textit{limited} control, since oftentimes, the input to the system, e.g. images, is captured by the device like surveillance camera, to which the adversary does not have direct access, making a lot of adversarial examples unrealistic.

To bridge the gap between the theoretic results and real-world constraints, limited effort has been made recently on \textit{practical adversarial learning}. As a prominent example, a recent study demonstrates that it is feasible to find adversarial examples with all the changes made around one's eyes, so a 3D-printed special glass frame can help one impersonate a different individual during FR-based authentication~\cite{sharif2016accessorize}. Along the similar line, another study reports the possibility to strategically perturb the images of stop signs, using a printed-out alternative or stickers to mislead a Deep Neural Network (DNN) to classifying the stop sign into a speed limit~\cite{evtimov2017robust}.

Despite the first steps these studies took, the attacks they propose, however, are still less practical. 3D-printed glasses are cool but also conspicuous, which could easily arouse suspicion. Printed signs and stickers only work on simple targets like stop signs. What less clear now is how similar techniques can be applied to generate realistic makeup to cheat FR systems. So far, how to induce recognition flaws in a realistic, less detectable way remains to be an open question.

\vspace{5pt}\noindent\textbf{Invisible mask attack}. In this paper, we present the first approach that makes it possible to apply automatically-identified, unique adversarial example to human face in an inconspicuous way, and completely invisible to human eyes. As a result, the adversary masquerading as someone else will be able to walk on the street, without any noticeable anomaly to other individuals but appearing to be a completely different person to the FR system behind surveillance cameras. This is achieved by using Infrared (IR), which cannot be seen by human eyes but can still be captured by most street cameras, surveillance devices and even smartphone cameras today. Given the tiny sizes of lit IR LEDs, smaller than a penny, we show that they can be easily embedded into a cap, and can also be hidden in an umbrella and possibly even hair or a wig. Once the devices are turned on, infrared dots will be projected on the strategic spots on the carrier's face, subtly altering her facial features to induce a misclassification in an FR system. This helps the attacker to evade detection or more seriously, through adjusting the positions, sizes and strengths of the dots, to impersonate a different person to pass FR-based authentication. The attack is called \textit{invisible mask attack} (IMA) in our research.

Launching an IMA is nontrivial. To find out how to strategically deploy LEDs and set their parameters, we have to seek adversarial examples that fit right in the shape, size, lightness and color of infrared prints which those devices are capable of generating. For this purpose, we developed an algorithm that searches adversarial examples for the attacker-victim pair in which the perturbations are the combination of dots that can be lit by infrared LEDs, making the adversarial examples implementable. Moreover, we designed a device that consists of three infrared LEDs on the peak of a cap, with which attacker can implement adversarial examples to attack real world systems. To help attackers implement the examples solved by our algorithm on the cap device, we designed a calibration tool with which attackers can adjust the LEDs on cap with ease.

In our experiment, with our developed algorithm and device, we successfully launched both dodging and impersonation attacks. The success rate for dodging is 100 percents and for the impersonation, the possibility of finding adversarial examples is over 70\%, according to our large scale study over LFW data set.


\vspace{5pt}\noindent\textbf{Contributions}. We summarize our contributions as follows:
\begin{itemize}
	\item \textit{Infrared-based stealthy facial morphing}. We come up with the first infrared based techniques capable of stealthily morphing one's facial features to avoid FR detection or impersonate a different individual.  Our study shows that this is completely feasible, even in the presence of a real-world mainstream FR systems. The outcomes of the research make an important step toward practical adversarial learning, showing that the attacks on DNN-powered FR systems are indeed realistic.

    \item \textit{New algorithm for finding actionable adversarial examples}.  We developed a new algorithm to search for adversarial examples under the constraints set by the limitations of commercial-off-the-shelf LEDs.  Our approach can find the attack instances satisfying the constraints with a high probability. We further show that such instances can be practically constructed through deploying the LEDs and adjusting their brightness and positions.

    \item \textit{Implementation and evaluation}. We implemented our attacks against FaceNet, a popular FR system, and demonstrated the effectiveness of our techniques against this real-world system. Besides, we launched a large scale study over LFW data set to demonstrate the possibilities of working out an adversarial example.
\end{itemize}

\section{Background}
\label{sec:bgd}

\noindent\textbf{Face embeddings}. Face embedding is the technique that utilizes a DNN to map a photo of a face into a vector. This allows a pair of face photos to be compared by calculating the distance between their vectors, as they are all in the shared vector space. Face embedding enables face authentication, face searching and face clustering, through thresholding the distance between photos, calculating a given photo's nearest neighbors and clustering a set of photos, based upon their vectors.
The most famous face embedding system is FaceNet~\cite{schroff2015facenet}, built by Schroff \etal  at Google in 2015, which achieved a 99.63\% accuracy on the LFW face data set.

\vspace{5pt}\noindent\textbf{Face authentication}. Using the face embedding technique, face authentication can be performed as follows. The authentication system firstly takes a photo of the user as its input, and then identifies the largest face in the photo, crop it out and fit it to a given size. The cropped image will then be centered according to the positions of eyes and lip detected using face land mark predictors. The centered image becomes an input to the face embedding DNN and a vector can then be generated and stored as a profile for the user.

Each time when a user is verified by the authentication system, the system takes a photo of the user and goes through the aforementioned steps to generate a vector. The similarity between the vector and the profile is then calculated to find out whether it exceeds a pre-defined threshold. The user passes the authentication if it does. Otherwise, he fails.

\vspace{5pt}\noindent\textbf{Infrared}. Infrared is the light that has a longer wavelength than the visible light. It can be produced by LEDs. There are two kinds of commonly seen IR LEDs available on the market, the 850nm and the 940nm. The IR emitted by both of them cannot be directly observed by humans.

On the other hand, IR can be captured by camera sensors. These sensors are built upon three types of units:  R, G and B, which are sensitive to Red, Green and Blue lights respectively. In the meantime, they can also detect other lights: e.g., Type B unit produces positive output even when illuminated by infrared, though most camera sensors do not have a specific infrared unit. As a result, camera sensors can generate an image very different from what people see, when the object in the image is exposed to infrared.

\begin{figure}
	\centering \includegraphics[width=0.38\textwidth]{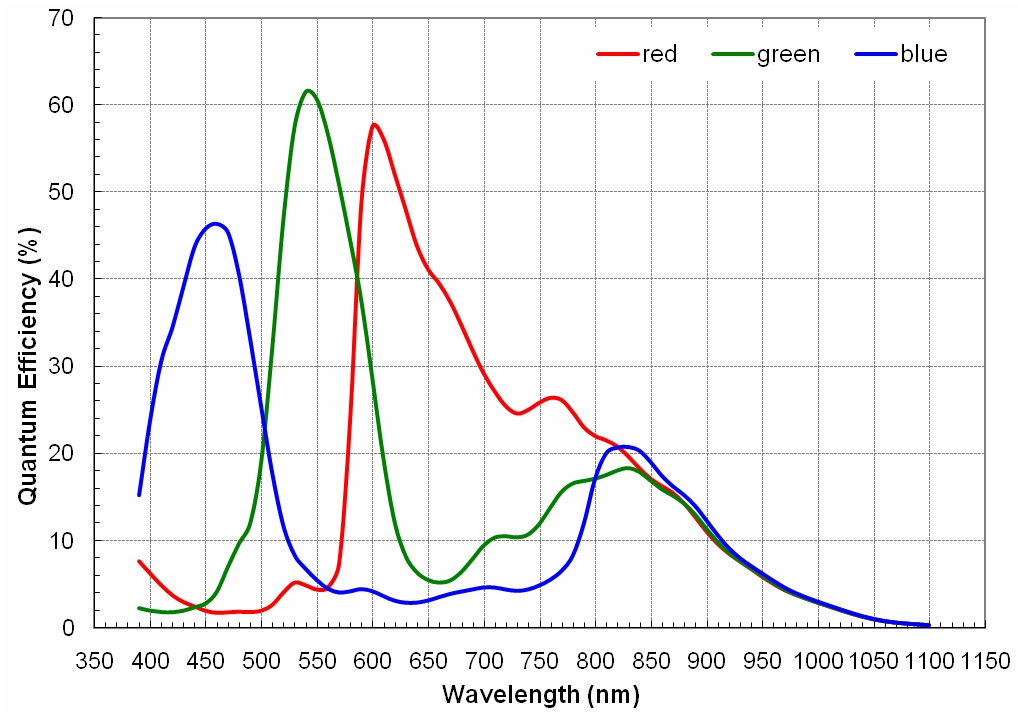}
    \caption{Frequency response curve of camera sensors.} \label{fig:ir_far}
\end{figure}

Fig.~\ref{fig:ir_far} shows the sensitivity of three types of units versus wavelengths (image acquired from Quora~\cite{ir_far}). As we can see, even for blue, which is the farthest away from the infrared range (at the 850nm for our LEDs), there still remains a considerable level of sensitivity. As a result, Type Blue unit can mistake infrared for blue light. Similarly, Red unit takes it for red and green unit mistakes it for green.

\vspace{5pt}\noindent\textbf{Adversarial examples}. Adversaries intentionally design some inputs to machine learning models, expecting the model's making mistakes. The inputs here are called adversarial examples.

While such an input is usually generated by adding a small perturbation to a legal input. The perturbation can be calculated according to the model, the legal input, and the target. Usually gradient optimizers solving equation~(\ref{eq:PerCal}) help adversaries calculate the perturbation, where $f(\cdot)$ is the learning model, $x$ is a legal input, $r$ is the perturbation, $y$ is the target output of the model and $J(\cdot,\cdot)$ is the loss function paired with the model.

\begin{equation}
\argmin_r J(f(x+r),y)
\label{eq:PerCal}
\end{equation}

Considering attacking a face authentication system with adversarial examples, the attacker can construct such an optimizer where $f(\cdot)$ is the face embedding model, $x$ is a photo of him, $y$ is the embedding (feature vector) of the victim that is generated with one of the victim's photo. $J(\cdot,\cdot)$ here is the loss function the model uses, which is usually a kind of distance between feature vectors, weighting the similarity of a pair of feature vectors. If only the optimizer figure out a loss below the threshold defined for the authentication system, the attacker can pass the authentication with the perturbation.

\section{Problem Statement}
\label{sec:pbm}
We consider two attacking scenarios that are likely to happen in our daily life. First, surveillance escaping that is an application of dodging technique and then, authentication cracking that utilizes impersonation.

\subsection{Dodging}
Thanks to the hi-tech devices being deployed around us, our cities are becoming safer than ever before. Long time ago, catching criminals was much more difficult and labor wasting. By comparison, today, with thousands of surveillance cameras distributed in cities, a criminal's each step can be recorded. Moreover, with abundant computational power, those targets of police can be found in a blinking of eyes, by just searching the criminals' faces among the video streams from the surveillance cameras.

However, a criminal may easily escape from such surveillance, if he understands the mechanisms behind the surveillance as well as some adversarial learning based dodging techniques like the one proposed in this work. Specifically, supposing he knows that the police men hunting for him use face recognition tools rather than \textit{manually} playback and examine every frame of every video captured by surveillance cameras, he only needs to use adversarial learning to fool the DNN models of the face recognition system to escape surveillance. As a consequence, in the video, he will not be recognized as himself, making the automatic searching null and void.

At the same time, the criminal does not want himself to be too distinctive, like warping his face with veil, otherwise people around him would rouse suspicion and pay more attention to him, at last, finding him a criminal.

Our proposed method aims at getting around the surveillance by wearing a cap mounting some button size LEDs on the peak, which is totally inconspicuous for nearby people.

\subsection{Impersonation}
The attacker may also want to impersonate a target victim, expecting to crack face authentication systems, in a real time and stealthily fashion.

Face authentication technique today is equipped on all kinds of mobile devices and security devices, like smart phones and laptops, door entrance or even boarder inspection, which are oftentimes targets of adversaries.

The attack must be in real time because usually, an attacker does not have much time to crack the authentication. For instance, in a library, a student want to just unlock someone's laptop and copy some files during when the owner goes to wash room, making attack methods that require fabricating dedicated device for each attack less practical.

The attack must also be stealthy, otherwise similar with dodging attacks, people around can notice the attack and rouse suspicion. The stealthiness is more crucial because, for scenarios where authentication is under manual supervision, guardians having noticed the attack will arrest the attacker. For example, at electronic boarder inspection channels that have been deployed by a lot of countries, users passing the channel that are supervised by officers.

\subsection{Threat Model}
In this work, unlike most previous works on adversarial learning, we consider the attack that can be practically implemented.

\zztitle{Target device integrity.} We consider that the integrity of the target devices has not been compromised during an attack: that is, the attacker does not have access to the device used for surveillance or authentication. Under this assumption, the adversarial examples presented by previous attacks, which require to change the input at pixel level precision, cannot be practically constructed, because such modification cannot be input to the device without compromising its integrity.

\zztitle{White box model.} The knowledge about the FR model is assumed to be publicly available, so the attacker can build an exactly the same model as the target one. We assume this because, in practice, application developers rarely train models by themselves. Instead, they tend to directly embed well-known models with pre-trained weights to the application or purchase models and weights from a specialized deep learning company. They do not train because 1) application developers typically do not have a large labeled dataset; without a huge amount of labeled data, the accuracy of the model cannot be guaranteed; 2) application developers usually do not have proper deep learning backgrounds to train the model.


\zztitle{Attacker's capability.} For authentication, the attacker is assumed to have already acquired at least one picture of his target. This can be done by, say, downloading from victim's social network or stealthily taking a photo of him.

For dodging and impersonation, the attacker is assumed to have a computer to launch the attack, a peak cap to mount the attack device that includes some button size LEDs, a battery and enough long wires.

\zztitle{Attacker's Goal.} For the dodging case, the attacker wants himself not to be recognized. More specifically, there are two ways to achieve this: one can fool the preprocessing network model, making attackers' face not detected; alternatively, she can make a detected face not recognized as the attacker, in which case the loss between the attacker's feature vectors with and without the dodging attempt should be larger than the threshold for classifying two vectors to the same person, as shown by the equation~(\ref{eq:sur}), where $x$ is the photo of the attacker, $(x+r)$ is the photo of the attacker with perturbation and $th$ is the aforementioned threshold.

\begin{equation}
 J(f(x+r), f(x) ) > th
\label{eq:sur}
\end{equation}

In the impersonation attack, the attacker wants himself to be identified as another (specific) individual, with a perturbation calculated by adversarial learning algorithms. To this end, the distance between an attacker launching the attack and his target should not exceed the threshold, as shown by the equation~(\ref{eq:auth}), where $y$ is the photo of the victim.

\begin{equation}
 J(f(x+r), f(y) ) < th
\label{eq:auth}
\end{equation}

\section{Dodging}
\label{sec:dod}
In this section, we explain how the attacker dodges surveillance cameras, by using the ``invisible mask'' --
our device issuing infrared for facial feature change.

\subsection{How Face Searching Works}
Before we explain how an attacker dodges surveillance, we introduce how face searching works.
\begin{itemize}
\item Video to be searched should first be split into frames, so that the searching can happen on images.

\item Each image is firstly preprocessed to extract its face portion only. In the preprocessing stage, images will be sent to a land marking model that identifies a set of land mark points for each face on the image, as shown by the equation~(\ref{eq:prep}).

\item For each set of land mark points, the face will be located and cropped out according to the positions of the land mark points, for later use.

\item Each face will be input to a face embedding model that converts it into a fixed length vector for future searching (e.g., using the k-NN algorithm).
\end{itemize}

\begin{equation}
landmark\_predictor(\cdot) \subset \{\emptyset, \{points\}^{n}\}
\label{eq:prep}
\end{equation}

\subsection{Design}

To dodge face searching, an attacker can either fail the land marking model or avoid the embedding model. 
We found that, with enough amount of infrared on face, the preprocessing step will fail, \ie, causing the face landmarking model to fail to output a set of valid land mark points, as shown in the equation~(\ref{eq:prep_fail}). Therefore we designed a device that can be mounted on a peak cap that emits enough infrared to fail the preprocessing step. So the attacker no longer needs to care about being identified.

\begin{equation}
landmark\_predictor(x) = \emptyset
\label{eq:prep_fail}
\end{equation}

\begin{figure}[ht]
	\centering \includegraphics[width=0.25\textwidth]{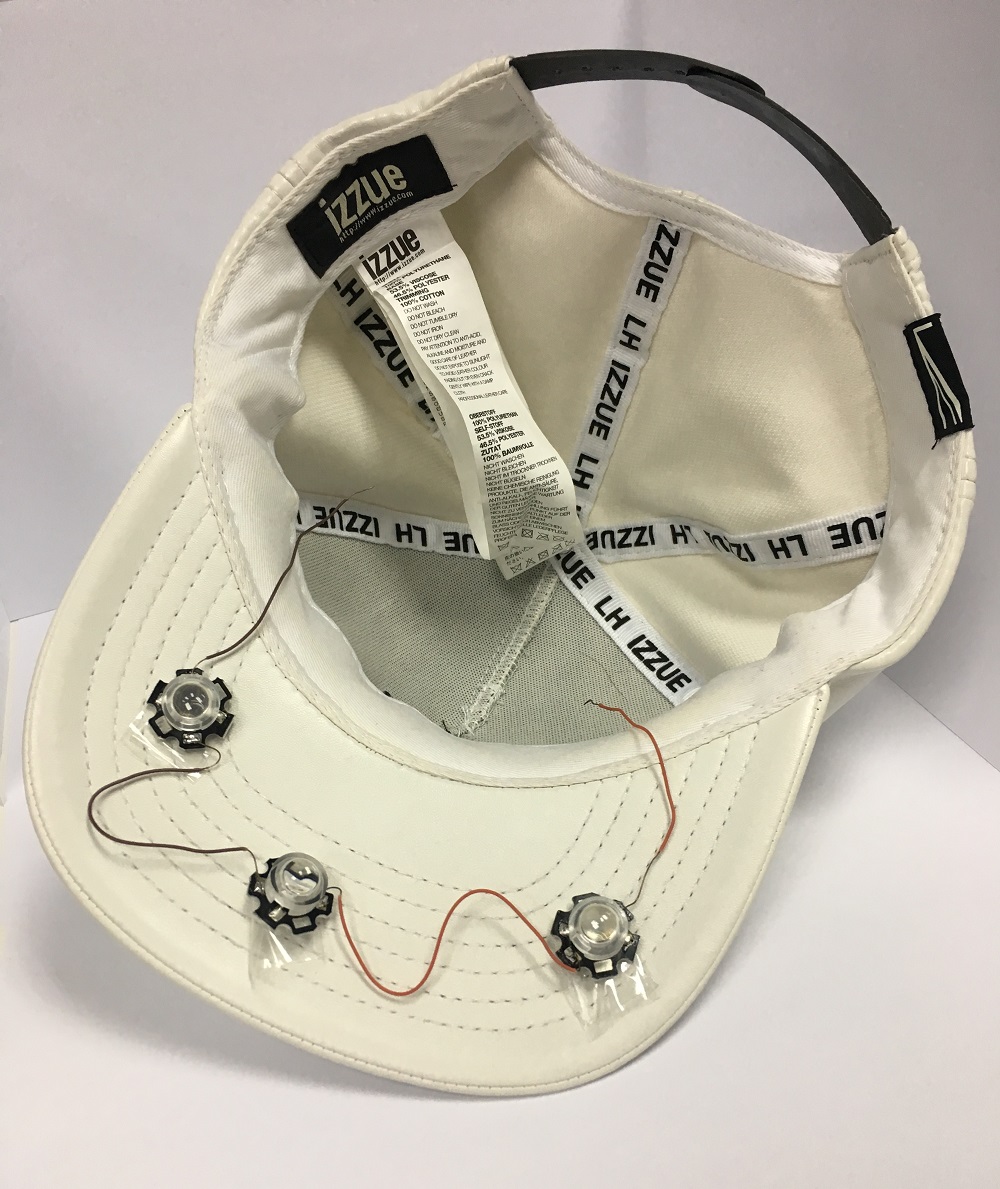}
    \caption{A cap mounting our device.} \label{fig:cap_1}
\end{figure}

\subsection{Device}

The dodging device consists of a peak cap, several LEDs and a battery. The LEDs are mounted on the peak of the cap and are facing the attacker's face. Each of the LEDs is used to interfere with the recognition of a zone of land marks. Fig.~\ref{fig:cap_1} shows a cap mounting our device.

\begin{figure}[ht]
	\centering \includegraphics[width=0.25\textwidth]{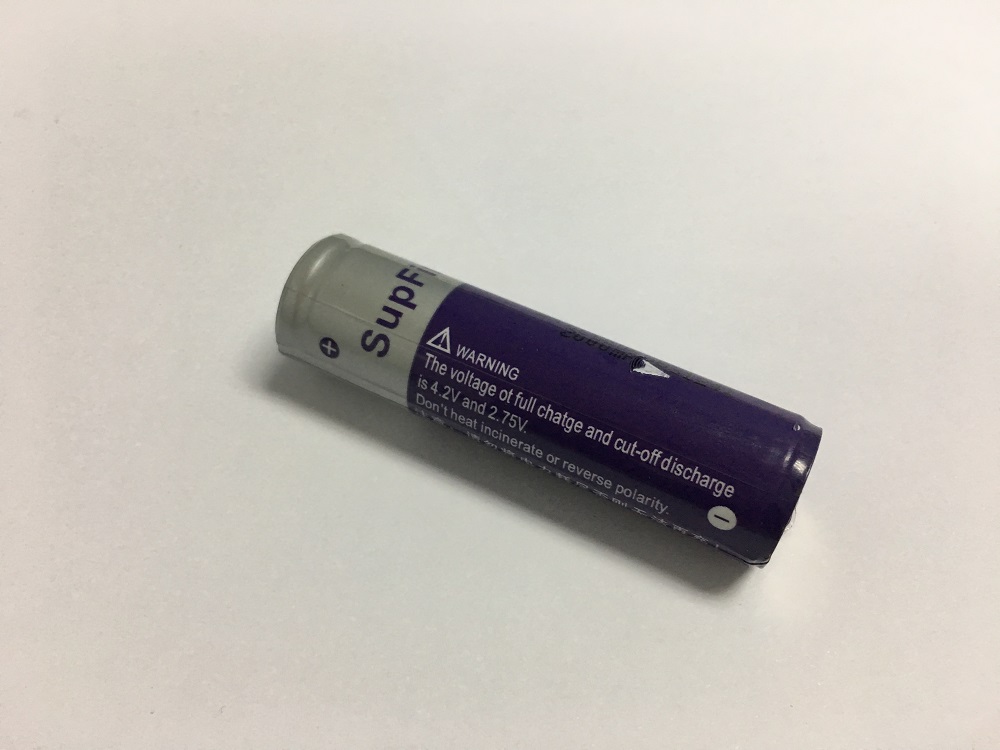}
    \caption{A 18650 battery powering the attack device.} \label{fig:battery}
\end{figure}

\zztitle{Power.} The device is powered by a 18650 battery, as shown in Fig.~\ref{fig:battery}. A single 18650 battery can support a LED for at least two hours, as we tested.

\begin{figure}[ht]
	\centering \includegraphics[width=0.25\textwidth]{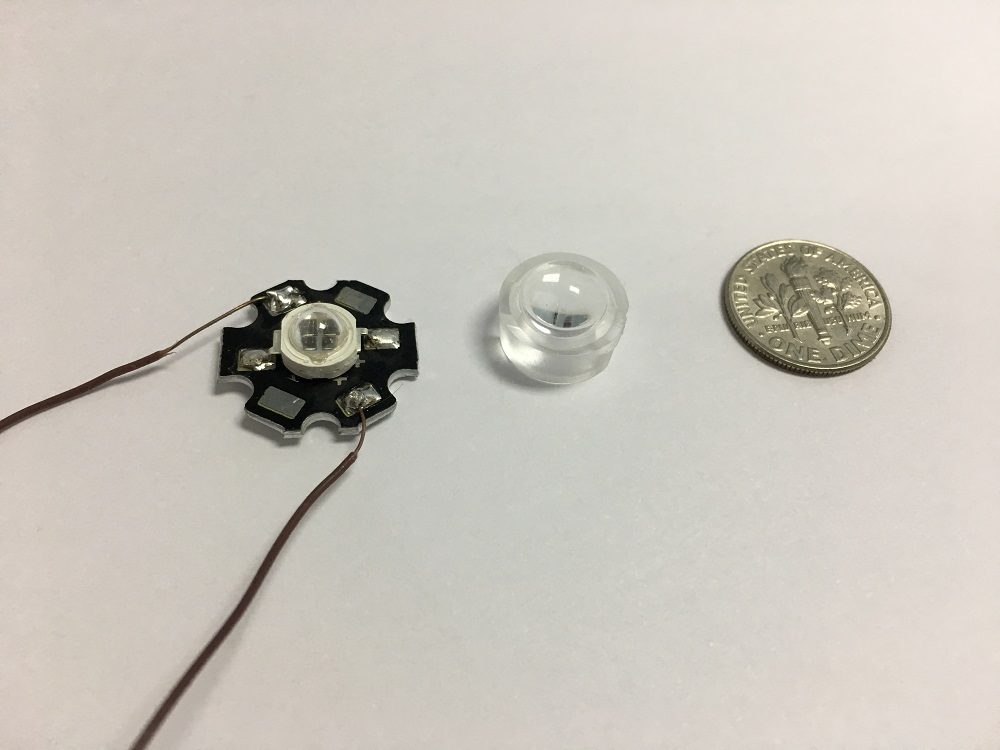}
    \caption{A LED as well as its lens.} \label{fig:led}
\end{figure}

\zztitle{Light source.} Each light module in the device is equipped with a 5W 850nm infrared LED as the light source (Fig.~\ref{fig:led}). 940nm LEDs can also be used here, but giving a darker illumination at the same given power consumption level.

Three LEDs are enough for causing the land mark collection step to fail, as our experiment demonstrates. If the attacker needs protection in brighter environments like outdoor, he may use more powerful LEDs and also mount more LEDs to his peak cap.

\subsection{Invisibility}
\begin{figure}[ht]
	\centering \includegraphics[width=0.25\textwidth]{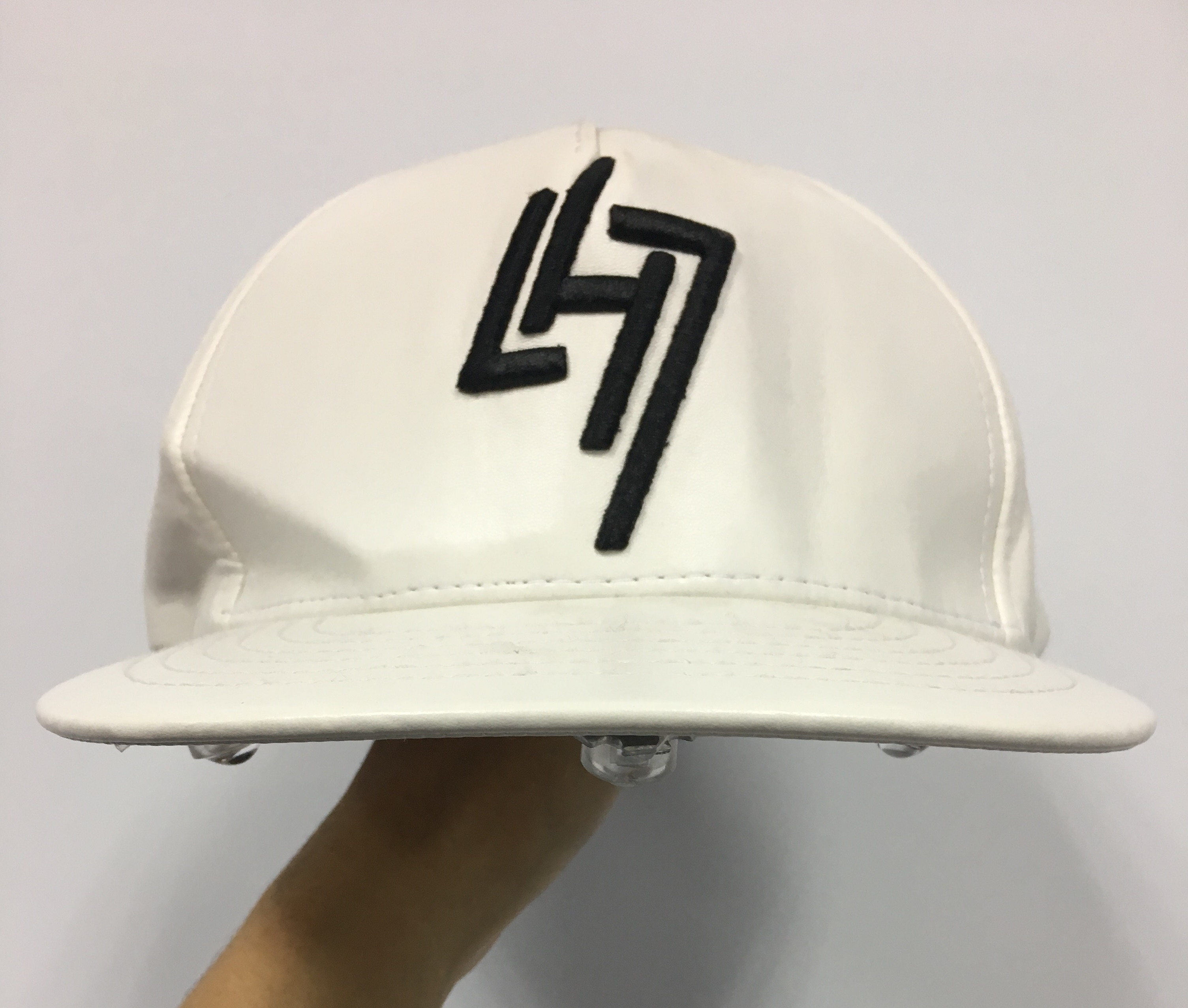}
    \caption{The device we devised to launch attacks.} \label{fig:cap_2}
\end{figure}

The device is overall  inconspicuous, as illustrated by Fig.~\ref{fig:cap_2}. Only the lenses could be seen by people nearby, which however are tiny and nearly transparent. The battery can be hidden inside the cap and the wires, if needed, can be hidden underneath the fabric of the peak. More importantly, the light emitted by the LEDs is totally invisible to human eyes.
\section{Impersonation}
\label{sec:imps}
In this section, we elaborate how the attacker finds an adversarial example that misleads the embedding model to recognize him as his victim target, by using only a pair of images of them. More importantly, how to implement such an adversarial example in real world.

\subsection{Overview}

Prior research that uses optimizers to find a perturbation does not apply to this scenario, because the solutions cannot be easily implemented using infrared LEDs. Specifically, an optimizer without special treatment does not restrict the solutions it finds to those that can possibly be realized using the combinations of infrared light spots.

In our research, we first construct a model that describes the infrared light spots on faces. Then, instead of optimizing the loss with adding pixel perturbations, we optimize the loss by adjusting light spots in line with our model on the attacker's photo. Each spot is produced by the model given a set layout parameters as its inputs. The models here shape the perturbation to ensure that it can be practically implemented using IR LEDs on the attacker's device, and the layout parameters determine the position, strength and size of the perturbation IR spot.


As a result, after the optimizer reaches the optimum through manipulating the parameters of all these models, the attacker obtains an image that is likely to be recognized by the embedding model as the victim, since the loss (that is, the distance) between the synthesized image and the victim's photo is minimized. The synthesized image after optimization is not only an adversarial example but also the one likely to be implemented by the adversary using the infrared LEDs.

\begin{equation}
\argmin_{r_i} J(f(x+\sum_i m(r_i),f(y))
\label{eq:auth_over}
\end{equation}

Equation~(\ref{eq:auth_over}) loosely describes the optimizer we use, where $m(\cdot)$ is the model for a light spot shown on the image, and $r_i$ is the layout parameters for a the light spot. $x+\sum_i m(r_i)$ after optimization is the adversarial example.

Given an adversarial example, the attacker can adjust the positions of LEDs on his attack device (Section~\ref{sec:dod}) and use the light spots generated by the device to approximate the perturbation observed from the example, with the help our calibration tool and several rounds of tuning.

Finally, the attack keeps the position of the devices and manages to impersonate the victim.

\subsection{Infrared Light Spot Modeling}

As mentioned earlier, a infrared light generated by an IR LED, though cannot be observed by human, produces a purple light spot on face that can be captured by camera sensors. We model such a light spot according to its position on face, brightness, size, but not color and shape.

\zztitle{Shape.} The shape of light spots varies, since the LED can be set to different angles when lit on the face. Also, human faces are not a flat plane, making light spots on faces hard to describe. To simplify the optimization process, we use circles, the closest shapes to the spots, to emulate the IR light's effects on faces.

Besides circle, when checking the brightness distribution of the light spot on face, we found that it attenuates from the center to its margin. We model this effect as a normally distributed attenuating dot.

\zztitle{Color.} The color is a little bit more difficult to decide. As mentioned before, cameras can sense infrared. However, the sensitivities of the three types of units toward infrared are different. Blue unit is the most sensitive one, while green type unit is the least sensitive one. As a result, camera will treat infrared as light purple, which is a mixture of more blue light and less green light.

To find out the accurate ratio between these three channels, we took two photos of a person with and without infrared light spot on his face respectively. Then we analyzed the difference between the two pictures at the infrared illuminated zones. The brightness values of the three types are of ratio 0.0852, 0.0533 and 0.1521 respectively. Therefore, when generating a light spot, we keep the ratio of the brightness values of the three channels for each pixel at 0.0852 : 0.0533 : 0.1521.

\zztitle{Size.} The size of a light spot is modeled as the standard deviation of normal distribution. The larger the standard deviation is, the larger the spot becomes. The standard deviation $\sigma$ here is a parameter that the optimizer can adjust.

\zztitle{Position.} The center coordinates of a light spot $\{p_x, p_y\}$ are also the input parameters of the optimizer.

\zztitle{Brightness.} The brightness values of different spots on the same face should be different, which is done by assigning an amplification coefficient $s$ to each spot.

In sum, the brightness of a pixel is the accumulation of the effects from all spots, while for each spot, the effect is decided by the amplification and the distance from the spot center to the pixel, which can be calculated by inputting the distance to the normal distribution probability density function (pdf), as shown by equation~(\ref{eq:brightness}). In this way, the center brightness of a spot is exactly the amplification coefficient $s$ and the spot size goes opposite with the standard deviation $\sigma$.

\begin{equation}
b(x,y) = s*norm\_pdf_{\sigma}((p_x - x)^2 + (p_y - y)^2)
\label{eq:brightness}
\end{equation}

\subsection{Synthesized Image}

As mentioned earlier, each light spot simulates the effect of the light from an LED. To get a face image with effects of all LEDs, we accumulate the effects of different LEDs together, and apply the result to the original image (the attacker's photo).


\begin{equation}
I_{syn} = I_{atk} + amp*coloring(\sum_i l_i)
\label{eq:synth}
\end{equation}

\begin{equation}
l_i =  \begin{bmatrix}
  b_i(1,1) &  ...   & b_i(1,w) \\
  \vdots & \ddots & \vdots \\
  b_i(h,1) &  ...   & b_i(h,w)
 \end{bmatrix}
\label{eq:bl}
\end{equation}

Equation~(\ref{eq:synth}) shows how the image is synthesized. $l_i$ here is the effect produced by one LED to the image by calculating the brightness for each pixel, shown by the equation~(\ref{eq:bl}). The function $coloring(\cdot)$ turns a gray scale image into a purple one using the fixed ratio between RGB as we tested. $amp$ here amplifies the effects and will also be added to the optimization variable list. To be noticed is that $amp$ here sets the strength of the perturbation, while $s$ makes brightness of spots uneven.

\subsection{Optimizer}
We choose an Adam optimizer to find adversarial examples. The objective function is the loss between the feature vector of the synthesized image and that of the victim's photo, as shown by equation~(\ref{eq:obj_func}). The optimization variable lists include the $amp$ and four aforementioned parameters for each light spot, \ie, ${\sigma,p_x,p_y,s}$ for each spot.

\begin{equation}
\argmin_{amp, \{\sigma,p_x,p_y,s\}_i} J(f(I_{syn}), f(I_{vtm}))
\label{eq:obj_func}
\end{equation}

By running the optimizer with his photo $I_{atk}$ and the victim's photo $I_{vtm}$, the attacker gets the minimum loss between them. If only the loss is no larger than the threshold of recognition, the $I_{syn}$ becomes an adversarial example. More importantly, the parameters yielding the minimum loss helps the attacker implement the example.

\subsection{Implementation}
Acquiring the adversarial example is the first step towards a success attack. The attacker also needs a device to implement the adversarial example. A light spot has three types of parameters that should all be adjustable with the device. For this purpose, we enhanced the dodging device with a PWM circuit and several variable angle lenses, for adjusting brightness and sizes respectively. Following we describe how we enhanced the device to support a given adversary example.


\zztitle{Positions.} ${(p_x,p_y)}$ describes the position of a light spot, which can be adjusted by choosing different mounting positions and angles for the LEDs. Once the attacker finds the correct position, he can fix it with tape or glue.

\begin{figure}[ht]
	\centering \includegraphics[width=0.25\textwidth]{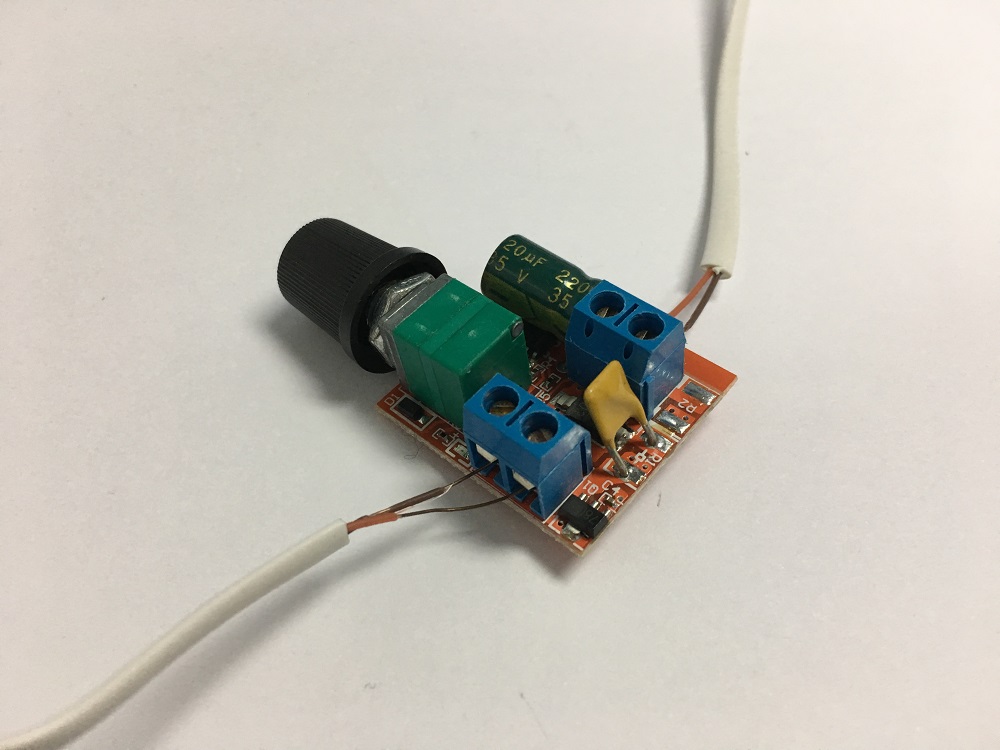}
    \caption{A PWM circuit to reduce the brightness of a LED.} \label{fig:pwm}
\end{figure}

\zztitle{Brightness.} The brightness of a light spot is adjusted through a newly introduced PWM circuit. It is because the input voltage of LEDs cannot be decreased to dim the bulb, instead, the brightness should be reduced through decreasing the time of power supply. Specifically, a PWM circuit turns on and off the LED at a very high frequency, say 10kHz. So, the brightness can be controlled by the ratio between the time intervals (0.1ms) in on state and off state (say 0.02ms and 0.08ms for a 0.1ms interval). Fig.~\ref{fig:pwm} shows a PWM circuit we use in our experiment.

\begin{figure}[ht]
	\centering \includegraphics[width=0.25\textwidth]{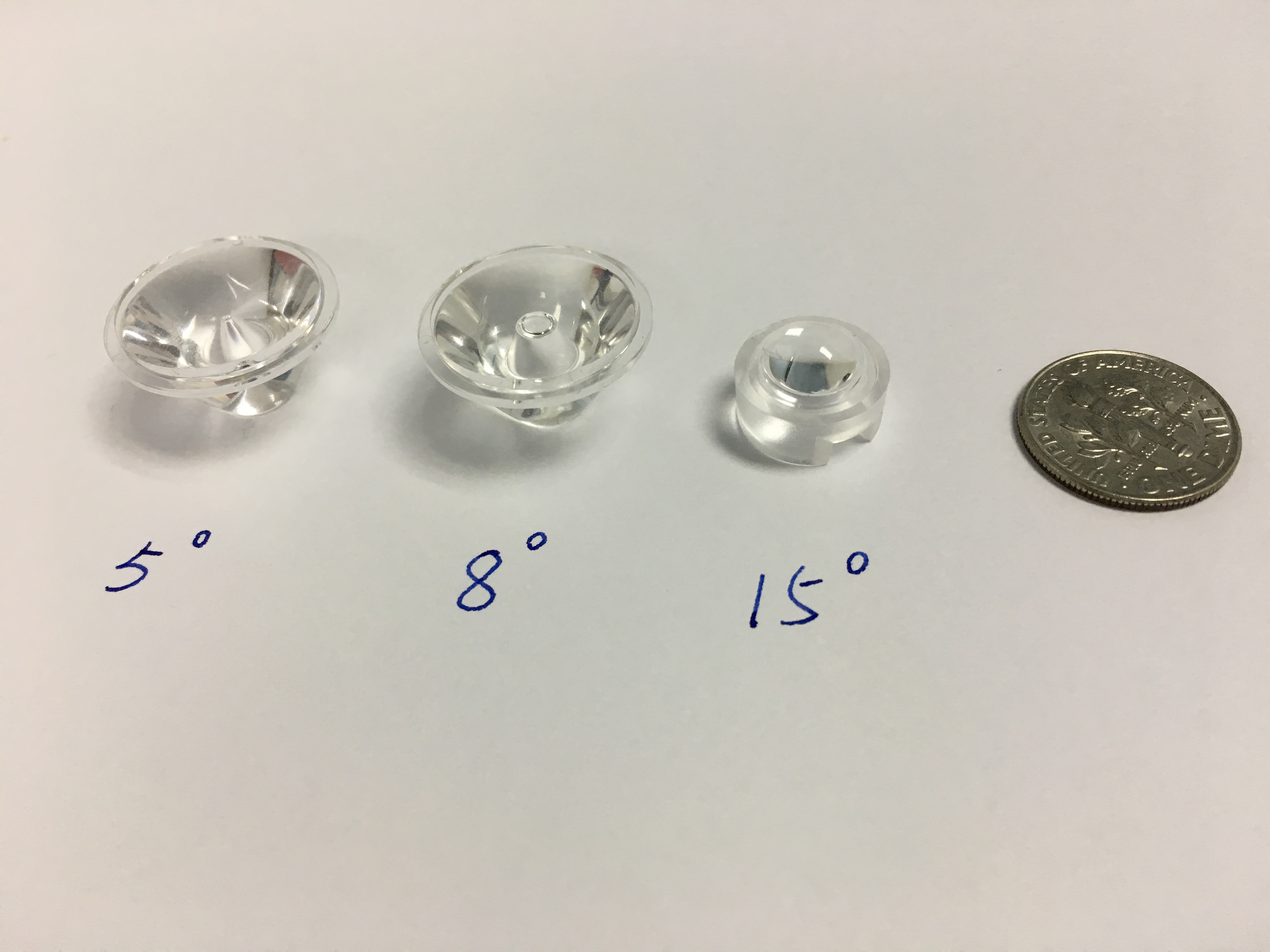}
    \caption{Three different kinds of lenses that can be used to adjust the size of the light spot.} \label{fig:lens}
\end{figure}

\zztitle{Size.} The size of a light spot can be adjusted through using different kinds of lenses, each of which has a different angle that results in different radius of the light spot. With the same angle and distance between a LED and an attacker's skin, a lens with wider angle makes the spot larger with less brightness, while a narrower yields a smaller but brighter one. Fig.~\ref{fig:lens} shows the lenses we used in our experiment.

\subsection{Calibration}
The attacker can use the interactive calibration tool we developed to make the light spots closer to the worked out adversarial example by instructing the attacker to tune the position, strength and size of each light spot. The tool works as follow:
\begin{enumerate}
	\item The attacker roughly adjusts the LEDs's position and chooses suitable lenses according to the adversarial example worked out by the optimizer for his target victim.
    \item Then he sits in front of a computer and starts the tool with the adversarial example as input. The tool asks the attacker to turn off all the LEDs and takes a photo (denoted as $I_{off}$) for the attacker.
    \item The tool asks the attacker to turn on the LEDs and takes a photo ($I_{on}$) for the attacker.
    \item Both photos are firstly preprocessed using alignment tool, after which only aligned faces left. The tool then calculates the difference between the two photos. \ie $I_{diff} = I_{on} - I_{off}$.
    \item \begin{itemize}
        \item The tool synthesizes each light spot according to the parameters of the adversarial example. Then it convolves each spot on $I_{diff}$ around the its theoretical center. Therefore the place where the maximum convolution value appears is the de facto center of light spot on $I_{diff}$, hence the center on $I_{on}$. The tool compares the center on $I_{on}$ and its theoretical place to know the offset and prompts the attacker the direction from the de facto center to the theoretical center (the direction of the offset).
        \item The tool calculates the average brightness of each light spot on $I_{diff}$ and divides it by the average brightness of $I_{on}$. The value is then compared with the theoretical value to see if the spot is too bright or too dim. The attacker will also know this.
        \item The tool searches around the light spot center to know where the brightness is reduced to half of the center. This value indicates the size of the spot. The value is also compared with the theoretical one to see if the light spot is too large or too small.
        \end{itemize}
    \item The attacker adjusts the positions, brightness and sizes of the LEDs according to the information provided by the tool. The tool shows the loss between $I_{on}$ and $I_{vtm}$ in a real time manner. So, the attacker can calibrate until he is satisfied.
    \item The tool refreshes and goes to the 3rd step to repeat the calibration process if it is not stopped by the attacker.
\end{enumerate}

\subsection{Fine Tuning}
After having finished the calibration by the tool, the attacker then fine tunes some spots to achieve a smaller loss by adjusting the LEDs along the direction toward decreasing the loss, with the real time loss feedback as reference, but without caring the parameters provided by the calculated adversarial example. This step can further improve success rate.

With this step, the deployed parameters of light spots actually deviates from the calculated ones. However it may yield a smaller loss, because the light spot model we used for optimization is a simplified model, which is not perfectly describing the actual spots. Therefore, the implementation with those parameters may not be local optimum in real case. Usually, the actual local optimum can be found by continue minimizing the loss with attackers' hands instead of an optimizer.

\section{Evaluation}
\label{sec:eval}
In this section, we introduce how the attack was evaluated, including some case studies of physical attacks and a large scale study showing the risk of the attack.

\begin{table}[ht]
\caption{Experiment environment used for the evaluation.}
\label{tab:env}
\centering
\begin{tabular}{c|c}
Face Embedding Model & Face Net \\ \hline
Model Version & 20170511-185253 \\ \hline
Preprocessing & CMU AlignDlib \\  \hline
Face Data Set & LFW
\end{tabular}
\end{table}

\subsection{Evaluation Environment}
Table.~\ref{tab:env} shows the models and data set we assume that our targets use. To be noticed is that we assume no proprietary model will be used. Instead, they use an embedding model pre-trained by well known organizations. Table.~\ref{tab:hardware} shows the platform we used to launch attacks.

\begin{table}[ht]
\caption{Experiment environment used for the evaluation.}
\label{tab:hardware}
\centering
\begin{tabular}{c|c}
CPU & Intel Core i7 7700T \\ \hline
Memory & 8G \\  \hline
OS & Ubuntu 16.04 Virtual Machine \\  \hline
Framework & Tensorflow 1.4
\end{tabular}
\end{table}

For the FaceNet embedding model, according to the paper\cite{schroff2015facenet}, squared L2 distance should be used to weight the distance between two feature vectors generated by their model. And they also gave thresholds for the LFW dataset, with which they got an excellent 99.63\% accuracy. We adopt the threshold 1.242 that was used in most cases by \cite{schroff2015facenet} over the LFW data set. Therefore, when a pair of faces has distance below the threshold, they will be recognized as from a same person, otherwise two distinct individuals.

\begin{equation}
p_{atk} = \frac{\eta p_{LED}}{\pi r^2}
\label{eq:ir_p}
\end{equation}

We didn't recruit a large number of volunteers to participate in our experiments, because we can not afford potential medical risks resulted from IR. Specifically, we roughly calculated the maximum radiation power of our LEDs on attackers' faces with the equation.(\ref{eq:ir_p}), where $p_{LED}$ is the power of an IR LED bulb, $\eta$ is the efficiency of LEDs and $r$ is the radius of a spot light. The estimation result is around 2100 $w/m^2$, which is 50\% more than the sun light (1413 $w/m^2$\cite{r_sun}). As we all know that someone staring at the sun for a period of time may suffer from sun burn, indicating that the IR LEDs may have similar effects. Therefore, we only do several case studies when implementing adversarial examples using our device. Besides, during the experiments, the experimenter had a 5 minutes cool down time between every 1 minute exposure to IR.

\subsection{Physical Dodging Attack}
We experimented how the device can help an attacker dodge surveillance. The experimenter was asked to wear the dodging device in front of a camera. The camera captures a clip of video of 5 seconds for the volunteer, during which the volunteer rotates his head as much as possible. The video was decomposed to frames and every frame was sent to the CMU's tool for preprocessing.

The results showed that no any single frame was recognized by the tool as a face appearing, indicating a successful dodging attack.

Yamada \etal proposed a similar dodging scheme with us\cite{yamada2012use}. The difference lies in that their device directly illuminates cameras, just like illuminating cameras using a flash light. The principles behind the two methods are totally different. Theirs are interfering the camera sensor while ours are fooling the alignment model. More importantly, our device is controllable so it can later be used to launch impersonation attack.

\subsection{Physical Impersonation Attack}
\ignore{
\begin{figure}[ht]
\centering
\includegraphics[width=0.25\textwidth]{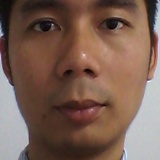}
\caption{The photo the attacker used to launch attack.}
\label{fig:attacker}
\end{figure}
}

\begin{figure*}[ht]
\centering
\begin{tabular}{m{5em} | c c c c}
Victim Name & Moby & Hoi-chang& Nan & Vladimir \\
\vspace{-60pt}Victim Photo
&\includegraphics[width=0.18\textwidth]{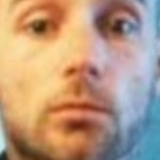}
&\includegraphics[width=0.18\textwidth]{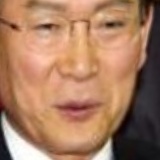}
&\includegraphics[width=0.18\textwidth]{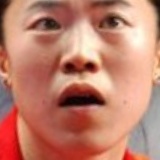}
&\includegraphics[width=0.18\textwidth]{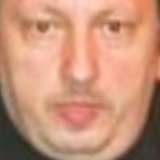} \\

\vspace{-60pt}Adversarial Example
&\includegraphics[width=0.18\textwidth]{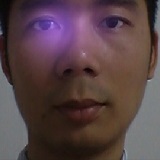}
&\includegraphics[width=0.18\textwidth]{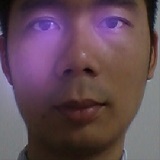}
&\includegraphics[width=0.18\textwidth]{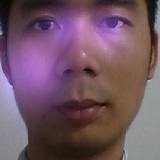}
&\includegraphics[width=0.18\textwidth]{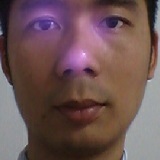} \\

\vspace{-60pt}Attacking Photo
&\includegraphics[width=0.18\textwidth]{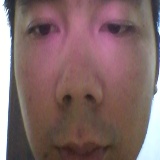}
&\includegraphics[width=0.18\textwidth]{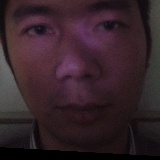}
&\includegraphics[width=0.18\textwidth]{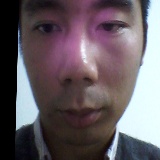}
&\includegraphics[width=0.18\textwidth]{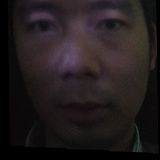} \\

\hline Original Distance & 1.36615 &1.32877 &1.33519 &1.27185 \\
\hline Theoretical Distance & 1.19221 & 1.12402 & 0.98804 & 0.94705\\
\hline Distance after Attack & 1.07773  & 1.12691 & 1.08065 & 1.23451
\end{tabular}
\caption{Physical impersonation attack result. Original distance means the distance between the embedding of the attacker and of the victim before launching attack. Theoretical distance means the distance between the calculated adversarial example and the victim. Distance after Attack means the distance between the victim and the attacker using our device after tuning.}
\label{fig:phy_ans}
\end{figure*}

To test if the calculated adversarial examples can be implemented in the real world, our experimenter implemented some examples using our attack algorithm and device.

\zztitle{Procedure.} For the impersonation attack, the volunteer played the attacker's role. And the experiment was conducted as follow:
\begin{itemize}
\item We took a photo for the experimenter.
\item The embedding of the photo was calculated. We also prepared a database of embeddings for all the photos in LFW dataset.
\item The attacker's embedding was searched over the database to get a subset of potential victims who have a distance that is not too high but above the threshold. The criterion was set to be below 1.4 but above 1.242 (the threshold).
\item We randomly selected four photos from the subset. They are assumed to be photos from four victims. For each of them, we ran the algorithm to search an adversarial example.
\item For each adversarial example, the volunteer used impersonation device and the calibration tool to implement the adversarial example. The time limit of tuning was set to 10 minutes.
\end{itemize}

During the experiment, we found the attacker need not to make full use of all the 5 LEDs. The optimizer takes 5 light spots as input and assumes them all variable. While when implementing an adversarial example, 3 LEDs were already enough for pulling down the distance to below threshold. Because, 1) the optimizer sometimes loses spots as some spots either were moved out of the face zone during optimization, or overlapped with each other. 2) 3 LEDs already produced a satisfying result and too much LEDs makes calibration and tuning difficult.

\zztitle{Experiment Result.} Fig.~\ref{fig:phy_ans} shows the result of the attack. As we can see, originally, the original distances are all above threshold, indicating an authentication system can recognize that it is not the victim in the photo. While our algorithm gave them all corresponding adversarial examples that theoretically can make distances fall below threshold. More importantly, the attacker can indeed implement those adversarial examples by using our device and consequently fool an authentication system.

\zztitle{Impact from Calibration.} As the experimenter summarized, the calibration played a really important role in approaching adversarial examples. Before calibration, the LED layouts by raw eye can hardly help him reduce the distance to below threshold. While with about 5 minutes' calibration, the distance can be largely reduced. But there was a case where it is still above threshold after calibration.

\zztitle{Free Tuning.} To get an even lower distance value, the attacker must tune the layout by his own. By emulating a ``manual optimizer", thus, adjusting the LEDs along the direction of pulling down the distance, an attacker can even get a lower distance than the theoretical example.

\newcounter{ii}
\begin{figure*}[t]
\centering
\forloop{ii}{1}{\value{ii} < 7}{
  \begin{subfigure}[t]{0.3\textwidth}
  	\centering
  	\includegraphics[width=\textwidth]{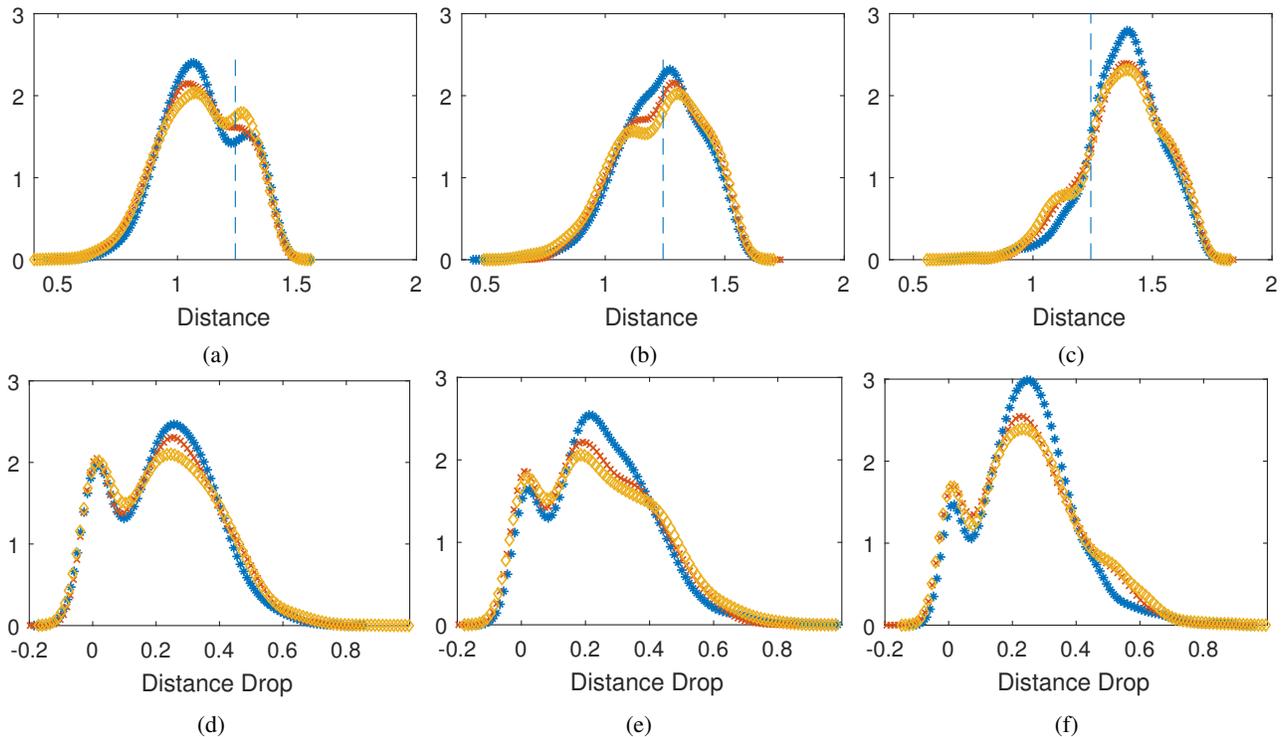}
    \caption{}
  \end{subfigure}
}
\caption{Large scale study result. Asterisk marked curves belong to the first attacker, cross marked curves belong to the second attacker and diamond marked curves the third. The two figures in the left column belong to the first group, the middle the second and the right the third. The figures in the first row show the PDF (probability density function) of the distances between the adversarial examples and the victims. The vertical line shows the threshold. The figures in the second row show the distance drops from using the attacker's original photos to using the adversarial examples.}
\label{fig:lg_res}
\end{figure*}

\subsection{Time Consumption and Complexity}
The time consumption of calculating an adversarial example mainly concentrates on the optimizer. In our experiment test bed, it takes 0.5 second for each iteration of the Adam optimizer. We restricted the number of iterations to 200. If the optimizer gives a distance lower than the threshold, 200 more iterations will be given to refine the adversarial example.

The complexity of each iteration is approximately equal to an epoch of training the model, as the variables newly introduced are only 21, which when comparing with the graph size of the embedding model is negligible. Besides, the time consumed for each iteration is also close to the time for a piece of embedding calculation.

\begin{figure}[h]
\centering
\includegraphics[width = 0.3\textwidth]{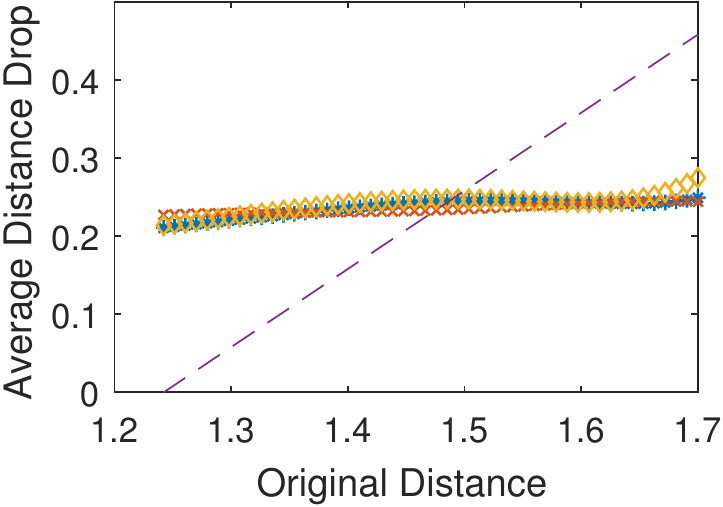}
\caption{Attacker's average distance gain with our algorithm versus their original distances. The dashed line signals the gain requirement for a successful attack.}
\label{fig:gain}
\end{figure}

\subsection{Large Scale Study}
Besides having physically implemented attacks, we also measured how likely the attacker can calculate a valid adversarial example when he has chosen a victim, through a large scale study covering the whole LFW data set.

\zztitle{Procedure.} In this study, we collected three photos from volunteers as three attackers' photos. Same as the prior part, photos from the dataset were used as victims's photos. For each attacker, we split photos in the dataset into five groups according to their distances from the attackers. Two groups were not tested, as their distances are either too small ( less than 1.242, model's false positive) or too large ( more than 1.7). We believe the attacker should ask someone else with a less distance from the victim to launch the attack if he finds the distance between victim and him too large.

The left three groups were experimented. Their distance ranges were (1.242, 1.4], (4.4, 1.55], (1.55,1.7] respectively. For each photo inside each group, we ran our algorithm to calculate an adversarial example for each attacker, and to see if the example helped the attacker to pull down the distance to below threshold.

\zztitle{Experiment Result.} Table.~\ref{tab:lg_res} shows the ratio of adversarial examples that made the distance below threshold and Fig.~\ref{fig:lg_res} shows the result of the large scale study in detail.

\begin{table}[ht]
\caption{Success rate for different victim groups with different original distances.}
\label{tab:lg_res}
\centering
\begin{tabular}{c c c c}
Original & Group 1 & Group 2 & Group 3 \\
Distance & 1.242 - 1.4 & 1.4 - 1.55 & 1.55 - 1.7 \\ \hline
Attacker 1 & 74.73\%  &  48.07\%  & 14.17\%  \\
Attacker 2 & 75.97\%  &  46.82\%  & 19.03\%   \\
Attacker 3 & 73.09\%  &  46.76\%  & 19.45\%
\end{tabular}
\end{table}

As we can see from Fig.~\ref{fig:lg_res}(a), for the group one who has the least original distance, attackers have a very high chance to work out a valid adversarial example. Comparing Fig.~\ref{fig:lg_res}(a) and Fig.~\ref{fig:lg_res}(b), Fig.~\ref{fig:lg_res}(c), we found the closer the original distance is, the more possible an attacker can pull the distance down to below threshold. Therefore, we imagine if there is a group of attackers instead of one single attacker, the success rate can be much higher, because it's more likely for a group of people to have at least one with close distance with their target victim. And the attackers can always deliberately elect the one who has similar skin color and other appearance features with the victim to launch the attack, which help the attacks to get a closer initial distance and thereby a closer result distance after optimization, and ultimately, a higher success rate.

When comparing Fig.~\ref{fig:lg_res}(d), Fig.~\ref{fig:lg_res}(e) and Fig.~\ref{fig:lg_res}(f), we found the algorithm is not sensitive to attackers, as they have very similar curves. The average distance drops are 0.2294, 0.2277 and 0.2302 for the three attackers respectively, as shown by Fig.~\ref{fig:gain}, and they measures attackers' benefit with our algorithm, also indicates the algorithm's insensitive toward attackers. According to the distance drop, we can see that attackers can expect a successful attack if the original distance is less than 1.45.

\section{Discussion}
\label{sec:disc}
\subsection{Limitations}
\zztitle{Not enough success rate.} As you may find from our evaluation result, the possibility of working out a valid adversarial example is not that high, comparing with the previous works. This is mainly because the degree of freedom attackers have in our work is much less than that in previous works. Specifically, in our work, a lot of graphics that may result in better adversarial examples are not in the searching range of optimizers, because of the limited number of LEDs an attacker can have and the fixed shape of light spots LEDs can produce. By comparison, in a lot of previous works, every pixel of the input, or a large chunk of pixels can be precisely manipulated by attackers, indicating a large amount of manipulatable variables or degree of freedom.

Nonetheless, we believe it is the limited amount of degree of freedom that makes the error tolerance good, when implementing adversarial examples.

\zztitle{Health concerns.} It is unclear if an attacker would get his eyes and face skins hurt, under the exposure of such a large amount of infrared in a long period of time. If this is true after confirmation from physicians, attackers may not dare to launch attacks with this method, worrying about their health.


\subsection{Future Work}

\zztitle{Automatic calibration.} It would be very good if the LEDs could be mounted on kinetic sockets that can be controlled by SoC (System on Chip), with which a mini program running on the SoC could dynamically calibrate the LEDs for an attacker, with the help of an embedding camera. All an attacker needs to do would be uploading victim's photo to the SoC, and the SoC calculates adversarial examples, adjusts and calibrates LEDs for the attacker.

\zztitle{Infrared projector.} Instead of using LEDs, attackers could devise a projector that can project any infrared graphical pattern. This device would be similar to a projector but with infrared LEDs as back light. In this case, the attacker need no longer restrict the model to a combination of spots. It could be nearly pixel level manipulatable images. With the help of this kind of device, the searching space would be largely extended, bring in a huge more amount of degree of freedom.

\zztitle{Defense.} Filtering out infrared at lens is not a feasible defense, though it is done by a lot of high end DSLRs. Specifically,  surveillance cameras need IR for imaging at night. While for authentication, filtering IR for ordinary lens is to expensive. We tested the IR sensitivity of the cameras on all the notebooks and mobile phones around us, including the front camera of iPhone 6s, MacBook Pro built-in camera and several top shipment devices. The results indicate that all of them are sensitive to IR.

Defenses against infrared are easy to implement, with adversarial training like method. Our calibration tool can already detect infrared spots on images. Therefore, images can be filtered before being sent to embedding models. And an image should be paid more attention if such infrared spots are found.

Recently, a lot of works aiming at defending adversarial learning attacks were proposed, but nearly all were turned to be useless~\cite{obfuscated-gradients} with a recent more advanced circumventing method. Defending against adversarial examples is still a open problem.

\zztitle{Black box extension.} Our work bases on a white box embedding model. However, recent work~\cite{sharif2016accessorize} showed that adversarial learning methods based on white box model can be migrated to black box model immediately, with the help of (PSO) Particle Swarm Optimization~\cite{eberhart1995new}, where the model structure is not required for optimization.

In fact, in our work, the embedding model was only used by attackers to calculate embeddings and by the optimizer to backwardly calculate gradients to optimize the loss. Distance calculations do not need the inside structure and weights of the model. Therefore white box assumptions are not used here. Attackers only need to query the black box to get the distance between a pair of images (synthesized and victim's), when there is black box only. For the gradient calculation, it is a little bit more complex. Intuitively, without knowing the expression of a function, approximate gradient values at given points can still be worked out, as shown by equation (\ref{eq:gradient}). Similarly, without the structure and weights of a model, the gradient of the model with respect to a layout variable can be worked out, though querying twice the black box learning model, \ie, the embedding of a current synthesized image and the embedding of a synthesized image with a little bit deviated layout variable.

\begin{equation}
 \frac{\partial f(x,y)}{\partial x}\bigg|_{x = x_0, y = y_0}  = \frac{f(x_0 + \Delta x,y_0) - f(x_0,y_0)}{\Delta x}
\label{eq:gradient}
\end{equation}

A recent work proposed a method to attack black box DNN\cite{narodytska2016simple}, during which they devised a local-search based technique to get a numeric approximation of a network gradient. Besides, there are other works~\cite{huang2017adversarial,hayes2017machine} can be implanted to help us to accommodate a black box model.

Admittedly, the approximated gradient has errors, comparing with the one worked out by back propagation. So it's unknown how much accuracy degradation will be introduced. 
\section{Related Works}
\label{sec:related}

\subsection{Practical Attacks to ML}
Sharif \etal proposed an accessory-based attack ~\cite{sharif2016accessorize}. They fool face recognition systems by special eyeglasses-frames. They designed a recursive algorithm to find printable adversarial examples, and launched attacks with adversarial example printed on eyeglasses. In their work, adversaries can wear the eyeglasses to dodge face recognition systems or impersonate a target victim. However, these special eyeglasses are so conspicuous, rousing suspicion from people around her/him. Besides, attackers need to fabricate device for each target. In contrast, wearing our device is inconspicuous and attackers could reuse the device for different targets.


Evtimov \etal ~\cite{evtimov2017robust} proposed a method to attack unmanned cars through adding perturbations to road signs. They pasted self made stickers on real road signs and successfully fooled recognition systems embedded in cars. Following their method, one can also paste stickers on face to fool the face recognition algorithm. However, unless if it is in Halloween nights, doing so would be so conspicuous. Therefore, we believe our approach is a more practical one.

Another kind of attack targeting face recognition is to merge a victim's facial characteristics into adversary's photos and present the synthesized photos in front of face authentication systems~\cite{li2014understanding}. However this kind of attack can hardly be stealthy, because it is a screen instead of a person in front of the authentication camera.

Zhang \etal proposed to use high frequency audio channel to inject command to fool voice assistant inside mobile phones or smart home devices~\cite{zhang2017dolphinattack}, which is also related to our work.

\subsection{Adversarial Learning}

Our approach relies on finding adversarial examples. Szegedy \etal \cite{szegedy2013intriguing} firstly defined adversarial examples and gave an insight of attacking visual recognition systems with adversarial examples. After that, plenty of works were proposed to improve the efficiency and robustness of the searching algorithm.

When searching adversarial perturbations, researchers mainly use three distance metrics, $L_0$, $L_2$, $L_{\infty}$. Minimizing different metrics results in different perturbations. Minimizing $L_0$, one gets perturbations with minimum number of pixels differing from those on the original input. The \textit{Jacobian-based Saliency Map} (JSMA)~\cite{papernot2016limitations} is such an attack optimizing $L_0$. By minimizing $L_2$, attackers obtain perturbations that have the minimum norm, in terms of Euclidean distance, across all pixels. Using this metric, Nguyen \etal ~\cite{nguyen2015deep} proposed an interesting attack that adds perturbations on a blank image to fool recognition systems.
Minimizing $L_{\infty}$, one finds perturbations with the smallest maximum-change to pixels. Under this metric, the algorithm manages to find a region of pixels with similar intensities to modify. An example of this kind of attack is \textit{Fast Gradient Sign Method} (FGSM)~\cite{goodfellow2015explaining}, which iteratively updates perturbations by stepping away a small stride along with the direction of the gradient.

More recently, Carlini \etal ~\cite{carlini2017towards} proposed an approach that significantly improved the efficiency of the searching algorithm. It demonstrated that using their approach can fool a number of defense techniques~\cite{carlini2017adversarial}. Besides, researchers~\cite{kurakin2016adversarial,hosseini2017blocking,madry2017towards} found the existence of transferability for perturbations, indicating that the perturbation generated for one system may be used to fool another system. Liu \etal ~\cite{liu2016delving} proposed an ensemble method to construct such transferable perturbations among different neural networks. Papernot \etal ~\cite{papernot2016transferability} demonstrated that transferability also exists between different machine learning techniques, such as Logistic Regression (LR), Support Vector Machine (SVM) and Nearest Neighbors (kNN).

\subsection{Security and Privacy of ML}
The security of Machine learning is attracting more and more attentions, Papernot \etal summarized recent findings in machine learning security and concluded an opposing relationship between the accuracy and security for machine learning algorithms\cite{papernot2016towards}.

\section{Conclusion}
\label{sec:con}
In this paper, we discovered that infrared can be used by attackers to either dodge or impersonate someone against machine learning systems. To prove the severeness, we developed an algorithm to search adversarial examples. Besides, we designed an inconspicuous device to implement those adversarial examples in the real world. As show cases, some photos were selected from the LFW data set as hypothetical victims. We successfully worked out adversarial examples and implemented those examples for those victims. What's more, we conducted a large scale study among the LFW data set, which showed that for a single attacker, over 70\% of the people could be successfully attacked, if they have some similarity.

Based on our findings and attacks, we conclude that face recognition techniques today are still far from secure and reliable when being applied to critical scenarios like authentication and surveillance. Researchers should pay more attention to the threaten from infrared.

\bibliographystyle{IEEEtran}
\bibliography{citation}

\end{document}